\documentclass{ptapap}

\author{Attila B\'odi}[Konkoly,SZTE]
\author{L\'aszl\'o Moln\'ar}[Konkoly]
\author{Emese Plachy}[Konkoly]
\author{R\'obert Szab\'o}[Konkoly]
\affil[Konkoly]{Konkoly Observatory, MTA CSFK, Budapest, Hungary}
\affil[SZTE]{Department of Experimental Physics and Astronomical Observatory, University of Szeged, Hungary}

\title{Shockwave Behaviour in RR~Lyrae Stars}

\begin{document}

\maketitle

\begin{abstract}

Here we present the detailed analysis of some modulated {\it Kepler} and {\it K2} RR~Lyrae stars that show peculiar bump progression in respect to the pulsation phase.

\end{abstract}

\section{Introduction}

Observations of RR~Lyrae stars in the {\it K2} mission revealed a peculiar bump progression in respect to the pulsation phase in some modulated RR~Lyrae stars. The apparent, unexpected shifts in the occurrence of the bumps (the signatures of shockwaves in pulsating stars) raise the question whether these objects are RR~Lyraes at all. We present the detailed analysis of light curves extracted by Extended Aperture Photometry (EAP; \citet{2017EPJWC.16004009P, 2012ascl.soft08004S})  to search for additional differences from the ``ordinary'' Blazhko stars. We found four stars in the {\it K2} fields and another one in the original {\it Kepler} frames that show modulations similar to those of V445\,Lyr \citep{2012MNRAS.424..649G}. The peculiar bumps always appear near the Blazhko minima, whereas the light curves become sinusoidal. These shockwaves propagate forward or backward with respect to the pulsation phase depending on the instantaneous pulsation phase at the time of emergence, respectively.

\section{Shockwave Behaviour of EPIC\,206072581}

\begin{figure}[h!]
  \centering
  \begin{minipage}{0.48\textwidth}
    \includegraphics[width=\textwidth]{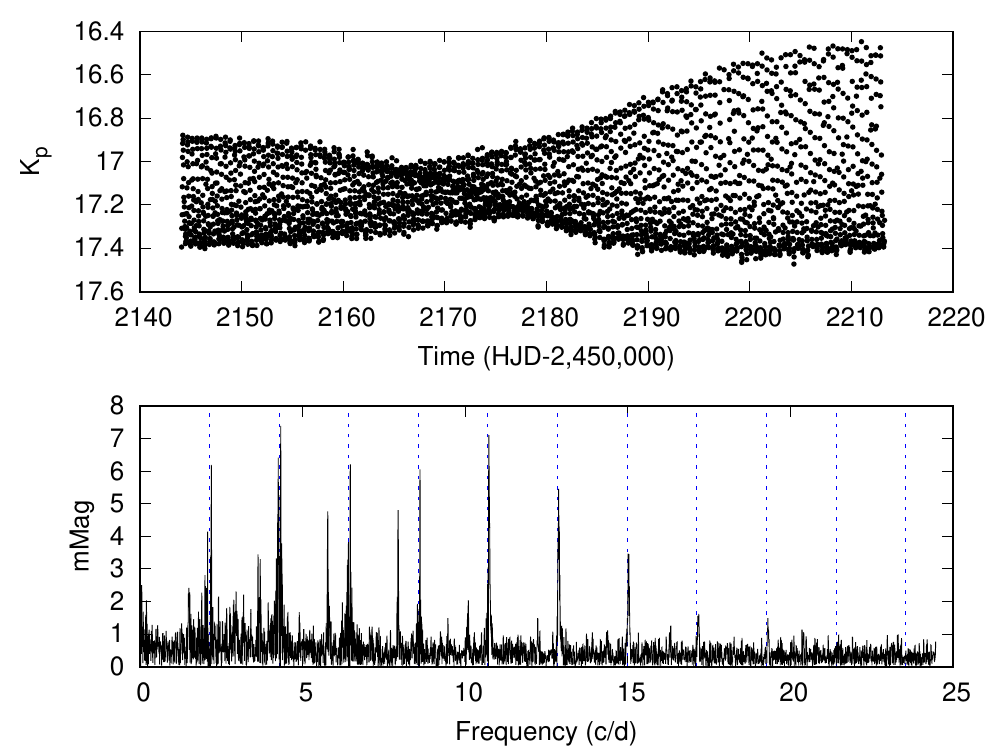}
    \caption{Light curve (top) and Fourier-spectrum (bottom) of EPIC\,206072581.}
    \label{fig:lc}
  \end{minipage}
  \quad
  \begin{minipage}{0.48\textwidth}
    \includegraphics[width=\textwidth]{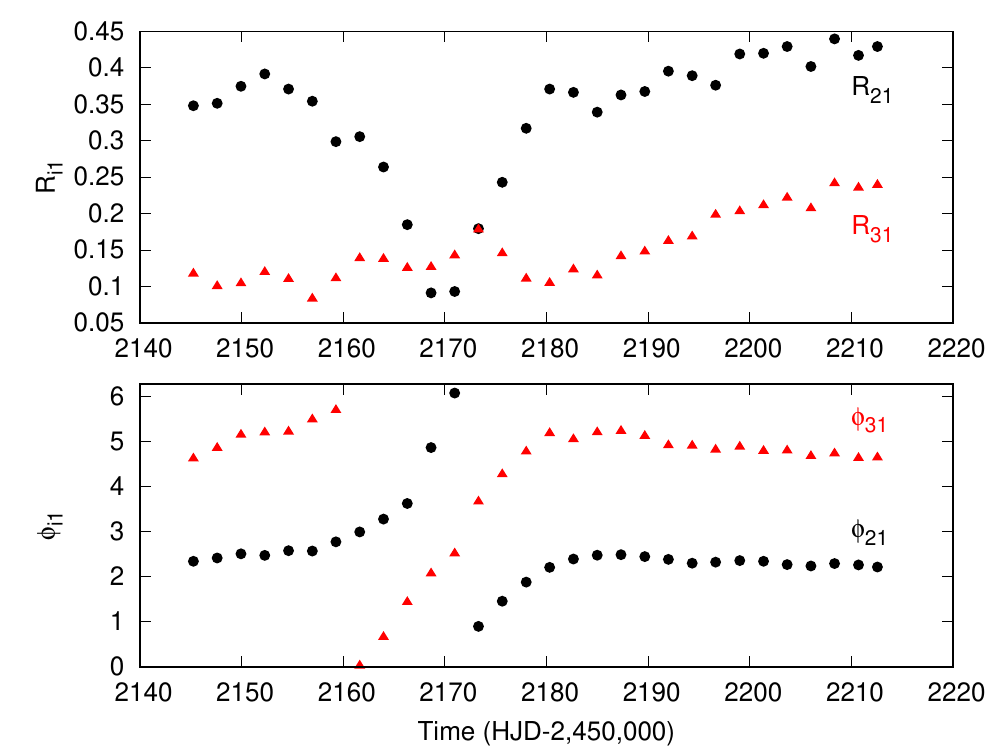}
    \caption{Time variation of the $R_{21}$, $R_{31}$ and $\phi_{21}$, $\phi_{31}$ Fourier-coefficients.}
    \label{fig:coeffs}
  \end{minipage}
\end{figure}

Fig. \ref{fig:lc} and \ref{fig:coeffs} show the light curve, Fourier spectrum, and the variation of the Fourier-coefficients, respectively, of EPIC\,206072581 as an example of the behaviour of the shockwaves.

In this case, the bump that appears at the maximum brightness of the
pulsation migrates downwards. During this, the $R_{21}$ amplitude
ratio decreases to $\sim$ 0.05, while the epoch-independent phase
differences do a counterclockwise loop (increasing phase
difference). In the Fourier spectrum, the amplitudes of some harmonics
of the secondary peak are higher than the one between $f_0$ and
$2f_0$. This can be a consequence of overlap with a higher frequency
variation. 

\section{Petersen Diagram}

\begin{figure}[h!]
\center
\includegraphics[width=9cm]{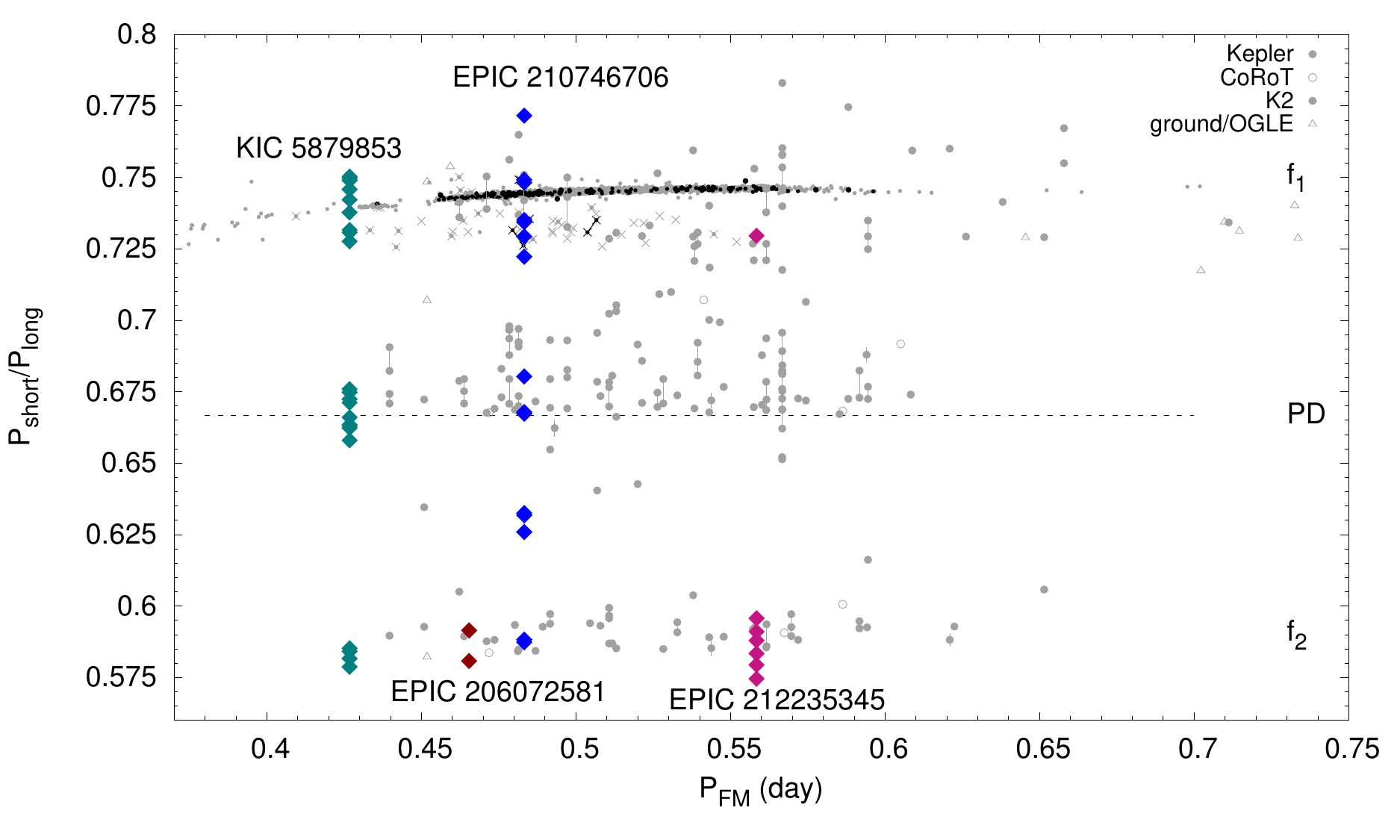}
\caption{The position of the peaks found between the fundamental mode and its first harmonics in the Petersen diagram.}
\label{fig:petersen}
\end{figure}

Additional peaks found between the fundamental modes and its first
harmonics are plotted in a Petersen diagram (adopted from \citet{2017EPJWC.16004008M}; Fig. 3.). Only two weird RR~Lyraes show the sign of period doubling, while the presence of the Blazhko modulation is unambiguous in the light curves. Three stars contain peaks around first overtone and four of them around the second overtone. Are all these excited modes radial or can they be non-radial as well?

\acknowledgements{Funding for the {\it Kepler} and {\it K2} missions are provided by the NASA Science Mission directorate. LM and EP are supported by the Bolyai J\'anos Research Scholarship of the HAS. This research received funding from the NKFIH grants K-115709, PD-116175, PD-121203, and from the Lend\"ulet LP2014-17 grant of the HAS.}

\bibliographystyle{ptapap}
\bibliography{bodi_shockwave_behaviour}

\end{document}